\documentclass[12pt]{article}
\usepackage{fancyhdr}

\usepackage{mathrsfs}
\usepackage[T1]{fontenc}
\usepackage{mathpazo}
\usepackage{setspace}
\usepackage{amsfonts}
\usepackage{amssymb}
\usepackage{amsmath}
\usepackage{epsfig}
\usepackage{latexsym}
\usepackage{color}
\usepackage{graphicx}
\usepackage{nicefrac}
\usepackage[latin1]{inputenc}
\usepackage{slashed}
\usepackage{multirow}
\usepackage{comment}
\usepackage{soul}
\usepackage{hyperref}
\usepackage[nosort]{cite}
\usepackage{datetime}

\usepackage[titletoc]{appendix}

\def\hybrid{\topmargin -20pt    \oddsidemargin 0pt
        \headheight 0pt \headsep 0pt
        \textwidth 6.25in       
        \textheight 9.25in       
        \marginparwidth .875in
        \parskip 5pt plus 1pt   \jot = 1.5ex}

\hybrid

\def\baselinestretch{1.2}

\catcode`\@=11

\def\marginnote#1{}
\newcount\hour
\newcount\minute
\newtoks\amorpm
\hour=\time\divide\hour by60
\minute=\time{\multiply\hour by60 \global\advance\minute by-\hour}
\edef\standardtime{{\ifnum\hour<12 \global\amorpm={am}%
        \else\global\amorpm={pm}\advance\hour by-12 \fi
        \ifnum\hour=0 \hour=12 \fi
        \number\hour:\ifnum\minute<10 0\fi\number\minute\the\amorpm}}
\edef\militarytime{\number\hour:\ifnum\minute<10 0\fi\number\minute}

\def\draftlabel#1{{\@bsphack\if@filesw {\let\thepage\relax
   \xdef\@gtempa{\write\@auxout{\string
      \newlabel{#1}{{\@currentlabel}{\thepage}}}}}\@gtempa
   \if@nobreak \ifvmode\nobreak\fi\fi\fi\@esphack}
        \gdef\@eqnlabel{#1}}
\def\@eqnlabel{}
\def\@vacuum{}
\def\draftmarginnote#1{\marginpar{\raggedright\scriptsize\tt#1}}

\def\draft{\oddsidemargin -.5truein
        \def\@oddfoot{\sl preliminary draft \hfil
        \rm\thepage\hfil\sl\today\quad\militarytime}
        \let\@evenfoot\@oddfoot \overfullrule 3pt
        \let\label=\draftlabel
        \let\marginnote=\draftmarginnote
   \def\@eqnnum{(\theequation)\rlap{\kern\marginparsep\tt\@eqnlabel}%
\global\let\@eqnlabel\@vacuum}  }

\def\preprint{\twocolumn\sloppy\flushbottom\parindent 2em
        \leftmargini 2em\leftmarginv .5em\leftmarginvi .5em
        \oddsidemargin -.5in    \evensidemargin -.5in
        \columnsep .4in \footheight 0pt
        \textwidth 10.in        \topmargin  -.4in
        \headheight 12pt \topskip .4in
        \textheight 6.9in \footskip 0pt
        \def\@oddhead{\thepage\hfil\addtocounter{page}{1}\thepage}
        \let\@evenhead\@oddhead \def\@oddfoot{} \def\@evenfoot{} }

\def\numberbysection{\@addtoreset{equation}{section}
        \def\theequation{\thesection.\arabic{equation}}}

\def\underline#1{\relax\ifmmode\@@underline#1\else
        $\@@underline{\hbox{#1}}$\relax\fi}

\def\titlepage{\@restonecolfalse\if@twocolumn\@restonecoltrue\onecolumn
     \else \newpage \fi \thispagestyle{empty}\c@page\z@
        \def\thefootnote{\fnsymbol{footnote}} }

\def\endtitlepage{\if@restonecol\twocolumn \else \newpage \fi
        \def\thefootnote{\arabic{footnote}}
        \setcounter{footnote}{0}}  

\catcode`@=12
\relax

\def\figcap{\section*{Figure Captions\markboth
        {FIGURECAPTIONS}{FIGURECAPTIONS}}\list
        {Figure \arabic{enumi}:\hfill}{\settowidth\labelwidth{Figure
999:}
        \leftmargin\labelwidth
        \advance\leftmargin\labelsep\usecounter{enumi}}}
 \relax
\def\tablecap{\section*{Table Captions\markboth
        {TABLECAPTIONS}{TABLECAPTIONS}}\list
        {Table \arabic{enumi}:\hfill}{\settowidth\labelwidth{Table
999:}
        \leftmargin\labelwidth
        \advance\leftmargin\labelsep\usecounter{enumi}}}
 \relax
\def\reflist{\section*{References\markboth
        {REFLIST}{REFLIST}}\list
        {[\arabic{enumi}]\hfill}{\settowidth\labelwidth{[999]}
        \leftmargin\labelwidth
        \advance\leftmargin\labelsep\usecounter{enumi}}}
 \relax
\makeatletter
\newcounter{pubctr}
\def\publist{\@ifnextchar[{\@publist}{\@@publist}}
\def\@publist[#1]{\list
        {[\arabic{pubctr}]\hfill}{\settowidth\labelwidth{[999]}
        \leftmargin\labelwidth
        \advance\leftmargin\labelsep
        \@nmbrlisttrue\def\@listctr{pubctr}
        \setcounter{pubctr}{#1}\addtocounter{pubctr}{-1}}}
\def\@@publist{\list
        {[\arabic{pubctr}]\hfill}{\settowidth\labelwidth{[999]}
        \leftmargin\labelwidth
        \advance\leftmargin\labelsep
        \@nmbrlisttrue\def\@listctr{pubctr}}}
 \relax
\makeatother
\newskip\humongous \humongous=0pt plus 1000pt minus 1000pt

\newif\ifdtup

\relax

\def\be{\begin{equation}}
\def\ee{\end{equation}}
\def\ba{\begin{eqnarray}}
\def\ea{\end{eqnarray}}

\def\del{\partial}

\def\a{\alpha}

\def\b{\beta}

\def\g{\gamma}

\def\d{\delta}

\def\m{\mu}

\def\l{\lambda}
\def\L{\Lambda}
\def\s{\sigma}

\def\cN{{\cal N}}

\def\no{\noindent}

\def\qq{\qquad}

\def\IR{\relax{\rm I\kern-.18em R}}

\def \ha {{1\over 2}}

\def \ov {\over}

\def\IR{\relax{\rm I\kern-.18em R}}
\def\IL{\relax{\rm I\kern-.18em L}}

\def\inv{^{\raise.15ex\hbox{${\scriptscriptstyle -}$}\kern-.05em 1}}

\def\Tr{{\rm Tr}}

\begin{document}

\renewcommand{\theequation}{\thesection.\arabic{equation}}
\csname @addtoreset\endcsname{equation}{section}

\newcommand{\beq}{\begin{equation}}
\newcommand{\eeq}[1]{\label{#1}\end{equation}}
\newcommand{\ber}{\begin{equation}}
\newcommand{\eer}[1]{\label{#1}\end{equation}}
\newcommand{\eqn}[1]{(\ref{#1})}
\begin{titlepage}
\begin{center}

\hfill CERN-TH-2018-017

${}$
\vskip .2 in

{\large\bf RG flows for $\lambda$-deformed CFTs}

\vskip 0.4in

{\bf E. Sagkrioti}$^1$,\ \ {\bf K. Sfetsos}$^1$\ \ and\ \ {\bf K. Siampos}$^2$
\vskip 0.1in

 {\em
${}^1$Department of Nuclear and Particle Physics,\\
Faculty of Physics, National and Kapodistrian University of Athens,\\
15784 Athens, Greece\\
}

\vskip 0.1in
 {\em
${}^2$Theoretical Physics Department, CERN, 1211 Geneva 23, Switzerland
}
\vskip 0.1in

{\footnotesize \texttt esagkrioti@phys.uoa.gr, ksfetsos@phys.uoa.gr, konstantinos.siampos@cern.ch}


\vskip .5in
\end{center}

\centerline{\bf Abstract}

\no
We study the renormalization group equations of the fully anisotropic $\lambda$-deformed CFTs involving the direct product of two current algebras at different levels $k_{1,2}$ for general semi-simple groups.
The exact, in the deformation parameters, $\beta$-function is found via the effective action of the quantum fluctuations around a classical background as well as from gravitational techniques. 
Furthermore, agreement with known results for symmetric couplings and/or for equal levels, 
is demonstrated. 
We study in detail the two coupling case arising by splitting the group into a subgroup and  the corresponding coset manifold which consistency requires to be either a symmetric-space one or a  
non-symmetric Einstein-space.

\vskip .4in
\noindent
\end{titlepage}
\vfill
\eject

\newpage

\tableofcontents

\noindent

\def\baselinestretch{1.2}
\baselineskip 20 pt
\noindent


\setcounter{equation}{0}

\section{Introduction}
\label{Introduction}

A class of integrable theories smoothly interpolating between exact CFTs in the UV and in the IR was constructed in \cite{Georgiou:2017jfi}.
These models are based on current bilinear deformations of two independent  WZW models at different positive levels $k_{1,2}$ and at the linear level they are of the form 
\begin{equation}
\label{Thirring}
S_{\l_1,\l_2}=S_{k_1}(\frak{g}_1)+S_{k_2}(\frak{g}_2)+\frac{\sqrt{k_1k_2}}{\pi}\int \text{d}^2\sigma \big((\l_1)_{ab}J^a_{1+}J^b_{2-}+(\l_2)_{ab}J^a_{2+}J^b_{1-}\big)+\cdots\ ,
\end{equation}
where $S_k(\frak{g})$ is the WZW action for a group element $\frak{g}\in G$ of dimension $\text{d}=\text{dim}G$
and the currents $J^a_\pm$ are given by
\begin{equation*}
J^a_+=-i\,\Tr(t_a\partial_+\frak{g}\frak{g}^{-1}), \quad J^a_-=-i\,\Tr(t_a\frak{g}^{-1}\partial_-\frak{g})\ ,
\end{equation*}
where the $t_a$'s are Hermitian matrices with $[t_a,t_b]=if_{abc}t_c$ and the structure constants $f_{abc}$ are real.
When a current has an index 1 or 2, this means that one should use the corresponding group element in its definition.

\no
Notice that the above models are driven away from the CFT point by mutual interaction of the currents of
the two independent WZW actions via current bilinears.  
This is different than the original $\l$-deformations introduced in \cite{Sfetsos:2013wia} in which the currents belongs to the same WZW action. 

\no   
Returning to \eqn{Thirring}, for finite coupling matrices $\l_{1,2}$ the action takes the form \cite{Georgiou:2017jfi}
\begin{equation}
\label{initialaction}
\begin{split}
&S_{\l_1,\l_2}=S_{k_1}(\frak{g}_1)+S_{k_2}(\frak{g}_2)+\\
&+\frac{1}{\pi}\int \text{d}^2\sigma\, \Tr\left\{\big(J_{1+}\phantom{0} J_{2+}\big) \binom{k_1\L_{21}\l_1D_2^T\l_2 \quad k_2\l_0\L_{21}\l_1}{k_1\l_0^{-1}\L_{12}\l_2\quad k_2\L_{12}\l_2D_1^T\l_1} \binom{J_{1-}}{J_{2-}}\right\}\,,
\end{split}
\end{equation}
generalizing the symmetric case \cite{Georgiou:2016urf}, where $D_{ab}=\Tr(t_a\frak{g}t_b\frak{g}^{-1})$ and
\begin{equation*}
\L_{12}=(\mathbb{I}-\l_2D_1^T\l_1D_2^T)^{-1},\quad \L_{21}=(\mathbb{I}-\l_1D_2^T\l_2D_1^T)^{-1}\,,
 \quad \l_0=\sqrt{\frac{k_1}{k_2}}\, .
\end{equation*}
We refer to \cite{Georgiou:2017jfi,Georgiou:2016urf},  for details of the derivation and further properties.

Following the lines of discussion in \cite{Georgiou:2017aei}, factorization of correlators involving current operators as well as composite current operators, implies that the $\beta$-function for the couplings $\l_1$ and $\l_2$ are the same as in the single $\l$-deformed theory, since the correlation functions from which they are derived involve only currents and as such they take the form of 
{\it two copies} of $\l$-deformed models. In particular, in the case of isotropic couplings,
i.e. $(\l_i)_{ab}=\l_i \d_{ab}$, these read \cite{Georgiou:2017jfi}
\begin{equation}
\label{dukjdodlk}
 \frac{\text{d}\l_i}{\text{d}t}=-\frac{c_G}{2\sqrt{k_1k_2}}\frac{\l_i^2(\l_i-\l_0)(\l_i-\l_0^{-1})}{(1-\l_i^2)^2}\ ,\qquad i=1,2\ .
\end{equation}
where $t=\ln\mu^2$ and $\mu$ is the energy scale 
and $c_G$ is the second Casimir in the adjoint representation defined from the relation $f_{acd}f_{bcd}=c_G\delta_{ab}$.

The goal of this work is to obtain the RG flows for generic couplings matrices
$(\l_{1,2})_{ab}$ in the action \eqref{initialaction} and to show that the result takes the form of \eqref{betadouble} below 
with the  definition \eqref{calN}.
The plan of this work is as follows: In section \ref{RGflows}, we tackle initially the single coupling matrix case using three independent methods, that is the one-loop effective theory for quantum 
fluctuations around a classical background, gravitational techniques and a CFT approach. Then, we work out the two coupling matrices case. In section \ref{RGflows.example}, we focus on an example based on a class of non-symmetric coset Einstein spaces.
We conclude with section \ref{concl}, where we summarize our results and we give an outlook on future directions.

\section{Computation of the RG flow equations}
\label{RGflows}

\subsection{The single coupling matrix}
\label{RGsingle}

In this section we will consider RG flows of the action \eqref{initialaction} when $\l_2=0$, while the other coupling matrix $\l_1$, renamed as $\l$, remains general. Then \eqn{initialaction} simplifies to 
\begin{equation}
\label{singlecase}
S_{\l}=S_{k_1}(\frak{g}_1)+S_{k_2}(\frak{g}_2)+\frac{\sqrt{k_1k_2}}{\pi}\int \text{d}^2\sigma\,\l_{ab}J_{1+}^aJ_{2-}^b\,.
\end{equation}
We shall compute its RG flows using three completely independent methods, the one-loop effective action for quantum fluctuations, gravitational techniques and CFT results,
all in agreement. Obviously, with this action, as compared to \eqref{initialaction}, the first two computational methods simplify considerably, especially the one involving gravitation techniques. In contrast, the CFT method is based on the form of the perturbation being bilinear in the currents and as such is insensitive to the details of the action for finite values of the couplings.

\subsubsection{The one-loop effective action}
\label{Appadu.general}

To compute the $\b$-function we need to specify a classical background solution and compute the quantum fluctuations around it. The discussion of this section goes along the lines of \cite{Appadu:2015nfa,Georgiou:2017aei}. In these works the 
method is described and 
applied for the isotropic case, i.e. when the $\l$'s are proportional to the identity which also correspond to 
integrable $\s$-models. However, until the present work it wasn't clear whether or not the method could be extendable to other
cases beyond integrable ones, let alone for general deformation matrices.

\no
The equations of motion of \eqref{singlecase} are given by \cite{Georgiou:2017jfi}
\be
\label{generaleom}
\begin{split}
&\l_0\partial_+A_--\l^{-T}\partial_-A_+=[\l^{-T}A_+,A_-]\,,\\
&\l^{-1}\partial_+A_--\l_0^{-1}\partial_-A_+=[A_+,\l^{-1}A_-]\,,
\end{split}
\ee
where
\be
\label{gaugefields}
A_+=i\l_0\l^T J_{1+}\,,\quad A_-=-i\l^{-1}_0\l J_{2-}\,.
\ee
At first we assume a background solution of \eqref{generaleom} for which the Lagrangian density is of course deduced from \eqn{singlecase} and reads
\be
\label{Lagrangianbackground}
{\cal L}^{(0)}={\cal L}_{k_1}(\frak{g}_1)+{\cal L}_{k_2}(\frak{g}_2)+\frac{\sqrt{k_1k_2}}{\pi} \l_{ab}J_{1+}^aJ_{2-}^b\ .
\ee
Next we vary the equations of motion \eqref{generaleom} obtaining the first order matrix equation
\begin{equation*}
\left(\begin{array}{cc}
\left(\l^{-T}\right)_{ab}\partial_- -i f_{acd}A_-^c\left(\l^{-T}\right)_{db} & -\l_0\d_{ab}\partial_+-i f_{abc}\left(\l^{-T}\right)_{cd}A_+^d 	\\
\l_0^{-1}\d_{ab}\partial_- + i f_{abc}\left(\l^{-1}\right)_{cd} A_-^d &-\left(\l^{-1}\right)_{ab}\partial_++i f_{acd}A_+^c\left(\l^{-1}\right)_{db} \end{array}\right)
\left(\begin{array}{c}
\d A^b_{+}\\
\d A^b_{-}
\end{array}\right)=0\,.
\end{equation*}
Then, the one-loop effective action in momentum space, after Wick rotating to Euclidean space and integrating out the fluctuations in the Gaussian path integral, reads
\be
\label{effectiveoneloop}
-{\cal L}^\text{eff}={\cal L}^{(0)}+\int^\mu\frac{\text{d}^2p}{(2\pi)^2}\ln\det{\cal D}^{-1/2}\,,\quad \text{d}^2p=\text{d}p_1\text{d}p_2\,,
\ee
where $\mu$ is the energy scale cutoff  and the matrix ${\cal D}$ is given by
\begin{equation}
{\cal D}=\left(\begin{array}{cc}
\left(\l^{-T}\right)_{ab}p_- -i f_{acd}A_-^c\left(\l^{-T}\right)_{db} & -\l_0\d_{ab}p_+-i f_{abc}\left(\l^{-T}\right)_{cd}A_+^d 	\\
\l_0^{-1}\d_{ab}p_- + i f_{abc}\left(\l^{-1}\right)_{cd} A_-^d &-\left(\l^{-1}\right)_{ab}p_++i f_{acd}A_+^c\left(\l^{-1}\right)_{db} \end{array}\right)\,,
\end{equation}
with $p_\pm=\ha \left(p_1\pm i p_2\right).$ To extract the $\beta$-function we need 
to compute using \eqref{effectiveoneloop} the logarithmic contribution in $\m$.  
To do so we first rewrite the determinant as
\begin{equation}
\label{jfklsnckskl}
\ln\det{\cal D}=\ln\det E+\Tr\ln\left(\mathbb{I}_{2\text{d}}+E^{-1}F\right)\,,
\end{equation}
where
\begin{equation}
\nonumber
\begin{split}
&E=\left(\begin{array}{cc}
\l^{-T} p_- & -\l_0p_+\mathbb{I}_\text{d} 	\\
\l_0^{-1}p_-\mathbb{I}_\text{d}  &-\l^{-1}p_+ \end{array}\right)\,,\quad F=\left(\begin{array}{cc}
 P & Q 	\\
 S & T \end{array}\right)\,,\\
 &P_{ab}=-i f_{acd}A_-^c\left(\l^{-T}\right)_{db}\,,\quad Q_{ab}=-i f_{abc}\left(\l^{-T}\right)_{cd}A_+^d\,,\\
 &S_{ab}=i  f_{abc}\left(\l^{-1}\right)_{cd} A_-^d\,,\quad T_{ab}=i f_{acd}A_+^c\left(\l^{-1}\right)_{db}\,.
\end{split}
\end{equation}
Then we compute the inverse of the matrix $E$ which we write as
\begin{equation}
\nonumber
E^{-1}=\left(\begin{array}{cc}
\frac{A}{p_-} & \frac{B}{p_-}	\\
\frac{C}{p_+} & \frac{D}{p_+} \end{array}\right)\,,
\end{equation}
where the various entries are given by
\be
\label{fospfks}
\begin{split}
&A=\l^T\tilde g^{-1}\,,\quad B=-\l_0\l^T\l g^{-1}\,,\quad C=\l_0^{-1}\l\l^T\tilde{g}^{-1}\,,\quad D=-\l g^{-1}\,,\\
&g=\mathbb{I}_\text{d}-\l^T\l\,,\quad \tilde{g}=\mathbb{I}_\text{d}-\l\l^T\quad\text{with}\quad \l g^{-1}=\tilde g^{-1}\l\,.
\end{split}
\ee
To proceed we expand the field dependent term of \eqref{jfklsnckskl}
\begin{equation*}
\Tr\ln\left(\mathbb{I}_{2\text{d}}+E^{-1}F\right)=-\frac{1}{p_+p_-}\Tr(AQ+BS)(CP+DR)+{\cal O}\left(\frac{1}{p_\pm^2}\right)\,,
\end{equation*}
which yields the only non-vanishing logarithmic contribution in \eqref{effectiveoneloop}.
It is simply a matter of algebra to prove that
\begin{equation*}
\begin{split}
&(AQ+BS)_{ab}=-i\l_0\big(\l^T\tilde g^{-1}\big)_{ac}\big(g\l^{-1}\big)_{db}\big(\l^{-1}\big)_{qp}{\cal N}_{cp}{}^d(\l,\l_0^{-1}) A_{+}^q\,,\\
&(CP+DR)_{ab}=-i\l_0^{-1}\big(\l g^{-1}\big)_{ac}\big(\tilde g\l^{-T}\big)_{db}\big(\l^{-1}\big)_{pq}{\cal N}_{cp}{}^d(\l^T,\l_0) A_{-}^q\,,
\end{split}
\end{equation*}
where we have defined
\be
\label{calN}
\mathcal{N}_{ab}{}^c(\l,\l^{-1}_0)=(\l_{ae}\l_{bd}f_{edf}-\l^{-1}_0\l_{ef}f_{abe})g^{fc}\quad\text{and}\quad g^{ab}=g^{-1}_{ab}\,.
\ee
Putting everything together in \eqref{effectiveoneloop} we find that
\be
\begin{split}
-{\cal L}_\text{eff}&={\cal L}^{(0)}-\frac{1}{2}\int^\mu\frac{\text{d}^2p}{(2\pi)^2}\frac{1}{p_+p_-}\mathcal{N}_{ac}{}^d(\l,\l_0^{-1})\mathcal{N}_{bd}{}^{c}(\l^T,\l_0)J_{1+}^aJ_{2-}^b\\
&=\cdots+\frac1\pi\left(\sqrt{k_1k_2} \l_{ab}-\ln\mu\,\mathcal{N}_{ac}{}^d(\l,\l_0^{-1})\mathcal{N}_{bd}{}^{c}(\l^T,\l_0)\right)J_{1+}^aJ_{2-}^b\,,
\end{split}
\ee
where the dots denote the two WZW actions which are $\l$-independent.
The one-loop $\b$-function is derived by demanding that the effective action is independent of the cutoff scale $\mu$, yielding
\begin{equation}
\boxed{
\frac{\text{d}\l_{ab}}{\text{d}t}=\frac{1}{2\sqrt{k_1k_2}}\mathcal{N}_{ac}{}^d(\l,\l_0^{-1})\mathcal{N}_{bd}{}^{c}(\l^T,\l_0)\,.
\label{beta}
}
\end{equation}
The levels $k_{1,2}$ retain their topological nature at one-loop in the large $k_{1,2}$ expansion
as they do not run with the energy scale.

The system of RG flows \eqref{beta} contains as a subclass the symmetric case $k_1=k_2$ \cite{Sfetsos:2014jfa,Georgiou:2017aei}. 
For an isotropic coupling matrix $\l_{ab}=\l\d_{ab}$ it is in agreement with \eqref{dukjdodlk}. Furthermore \eqref{beta} 
is invariant under the transformation\footnote{
\label{analyticks}
This transformation should be implemented as an analytic continuation: $k_{1,2} \to \text{e}^{i\pi} k_{2,1}$.
}
\begin{equation*}
\l\to\l^{-1}\ , \quad  k_{1,2}\to-k_{2,1} \ ,
\end{equation*}
due to the property
\begin{equation*}
{\cal N}_{ab}{}^c(\l,\l_0^{-1})\to\l_0\l^{-1}_{ad}\l^{-1}_{be}\l_{cm}{\cal N}_{de}{}^m(\l,\l_0^{-1})\,,
\end{equation*}
and also retains its form under the transformation
\begin{equation*}
\l\to\l^T\ ,\quad k_{1,2}\to k_{2,1}\ .
\end{equation*}
In fact a symmetric coupling matrix $\l$ remains symmetric under the RG flows, as it can be readily checked from \eqref{beta}.
For this class of coupling matrices there exists an analogue formula in the literature \cite{LeClair:2001yp}.  We show in section \ref{CFT.connection} that it is equivalent with \eqref{beta}.

\subsubsection{Gravitational techniques}
\label{RG.gravity}

We are going to re-derive \eqref{beta}, using gravitational methods along the lines of \cite{Sfetsos:2014jfa,Georgiou:2017aei}.
The line element of \eqref{singlecase} reads
\begin{equation}
\text{d}s^2=R^aR^a+\l_0^{-2}L^{\hat{a}} L^{\hat{a}} +2\l_0^{-1}\l_{ab}R^aL^{\hat{b}}\,,
\end{equation}
where
\begin{equation*}
\begin{split}
R^a=-i\, \Tr(t_a\text{d}\frak{g}_1\frak{g}_1^{-1}),\qq L^{\hat{a}}=-i\, \Tr(t_a\frak{g}_2^{-1}\text{d}\frak{g}_2)\,,\\
\text{d}R^a=-\frac{1}{2}f_{abc}R^{{b}}\wedge R^{{c}},\qq \text{d}L^{\hat{a}}=\frac{1}{2}f_{abc}L^{\hat{b}}\wedge L^{\hat{c}}\,.
\end{split}
\end{equation*}
Hence, the unhatted and hatted indices denote the Maurer--Cartan forms of $\frak{g}_1$ and $\frak{g}_2$ respectively. By introducing the vielbeins
\begin{equation}
\text{e}^a=R^a,\quad \text{e}^{\hat{a}}=\l_{ba}R^b+\l_0^{-1}L^{\hat{a}},  \label{veilbeins}
\end{equation}
as well as the double index notation $A=(a,\hat{a})$, the line element can be written as
\begin{equation*}
\text{d}s^2=\tilde{g}_{ab}\text{e}^a\text{e}^b+\text{e}^{\hat{a}}\text{e}^{\hat{a}}=G_{AB}\text{e}^A\text{e}^B\,.
\end{equation*}
Note that in the previous calculations, as well as in the following ones an overall 
factor of $\frac{k_1}{2\pi}$ is not included.
We may now proceed with the computation for the spin connection $\omega_{AB}$. Since the tangent metric $G_{AB}$ is constant, $\omega_{AB}$ is antisymmetric. A practical way to compute it is by first define the quantities $C^A{}_{BC}=-C^A{}_{CB}$ from
\begin{equation*}
\text{d}\text{e}^A=\frac{1}{2}C^A{}_{BC}\text{e}^B\wedge \text{e}^C,\quad C_{ABC}=G_{AD}C^D{}_{BC}\,.
\end{equation*}
Then simply
\begin{equation*}
\omega_{AB}=\omega_{AB|C}\text{e}^C=\frac{1}{2}(C_{ABC}-C_{CAB}+C_{BCA})\text{e}^C\,,
\end{equation*}
from which we can also extract the useful quantity $\omega_{AB|C}$. Employing the above along with (\ref{veilbeins}) we find that
\begin{eqnarray}
&&\omega_{ab}=-\frac{1}{2}(\tilde{g}_{ad}f_{dbc}-\tilde{g}_{cd}f_{dab}+\tilde{g}_{bd}f_{dca})\text{e}^c+\frac{1}{2}(\l_{dc}f_{dab}-\l_0\l_{ad}\l_{be}f_{cde})\text{e}^{\hat{c}}\,,\nonumber\\
&&\omega_{\hat{a}b}=\frac{1}{2}(\l_0f_{ade}\l_{bd}\l_{ce}-\l_{da}f_{dbc})\text{e}^c\,,\\
&&\omega_{\hat{a}\hat{b}}=-\l_0\l_{cd}f_{abd}\text{e}^c+\frac{1}{2}\l_0f_{abc}\text{e}^{\hat{c}}\,. \nonumber
\label{omega}
\end{eqnarray}
In order to find the torsion-full spin connections we need the two-form of \eqref{singlecase} which is given by
\begin{equation}
\label{B2form}
B=B_0+\l_0^{-1}\l_{ab}R^a\wedge L^{\hat{b}}\,,
\end{equation}
where $B_0$ is the two-form corresponding to the two WZW models with
\begin{equation*}
H_0=\text{d}B_0=-\frac{1}{6}f_{abc}\,R^a\wedge R^b\wedge R^c-\frac{\l_0^{-2}}{6}f_{abc}\,L^{\hat{a}}\wedge L^{\hat{b}}\wedge L^{\hat{c}}\,.\end{equation*}
The field strength of the two-form $B$ is
\begin{equation}
\begin{split}
H=\text{d}B=&-\frac{1}{6}\big(f_{abc}-3f_{abd}(\l\l^T)_{cd}+2\l_0\l_{ad}\l_{be}\l_{cf}f_{def}   \big)\text{e}^a\wedge \text{e}^b\wedge \text{e}^c+\\
&+\frac{1}{2}\big(\l_0\l_{ce}\l_{bd}f_{ade}-\l_{da}f_{dbc}   \big)\text{e}^{\hat{a}}\wedge \text{e}^b\wedge \text{e}^c
-\frac{\l_0}{6}f_{abc}\text{e}^{\hat{a}}\wedge \text{e}^{\hat{b}}\wedge \text{e}^{\hat{c}}\,. \label{H}
\end{split}
\end{equation}
The torsion-full spin connections are defined as
\begin{equation*}
\omega^\pm_{AB}=\omega_{AB}\pm\frac{1}{2}H_{ABC}\text{e}^C=\omega^\pm_{AB|C}\text{e}^C\,.
\end{equation*}
Using the above along with (\ref{omega}) and (\ref{H}) we find that
\begin{eqnarray}
&&\omega^+_{ab}=\left(-f_{abc}-\l_0\l_{ad}\l_{be}\l_{cf}f_{def}+(\l\l^T)_{ad}f_{dbc}+(\l\l^T)_{bd}f_{adc}\right)\text{e}^c\,,\nonumber\\
&&\omega^+_{\hat{a}b}=\big(\l_0\l_{bd}\l_{ce}f_{ade}-\l_{da}f_{dbc}\big)\text{e}^c\,,\\
&&\omega^+_{\hat{a}\hat{b}}=-\l_0f_{abd}\l_{cd}\text{e}^c \label{omega+}\,,\nonumber
\end{eqnarray}
and that
\begin{eqnarray}
&&\omega^-_{ab}=\big(\l_0\l_{ad}\l_{be}\l_{cf}f_{def}-(\l\l^T)_{cd}f_{dab}\big)\text{e}^c+\big(\l_{dc}f_{dab}-\l_0\l_{ad}\l_{be}f_{dec}\big)\text{e}^{\hat{c}}\,,\nonumber\\
&&\omega^-_{\hat{a}b}=0\,,\\
&&\omega^-_{\hat{a}\hat{b}}=-\l_0f_{abd}\l_{cd}\text{e}^c+\l_0f_{abc}\text{e}^{\hat{c}}\,.\nonumber
\label{omega-}
\end{eqnarray}
We are now in position to compute the torsion-full Ricci tensor by a rewriting
\begin{equation*}
R^{\pm}_{AB}=\partial_{C}\omega^{\pm C}{}_{A|B}-\omega^{\pm}{}_{AC|D}\omega^{\mp D|C}_B-\nabla_{B}^{\pm}\omega^{\pm C}{}_{A|C}\,.
\end{equation*}
The one-loop RG flow equations read \cite{honer,Friedan:1980jf,Curtright:1984dz}
\begin{equation}
\frac{\text{d}}{\text{d}t}(G_{MN}+B_{MN})=R^{-}_{MN}+\nabla^{+}_N\xi_{M}\ ,
\end{equation}
or equivalently in the tangent frame $\text{e}^A=\text{e}^A{}_M\text{d}X^M$
they take the form 
\begin{equation}
\label{oneloopRG}
\frac{\text{d}}{\text{d}t}(G_{MN}+B_{MN})=(R^{-}_{AB}+\nabla^{-}_B\xi_{A})\text{e}^A{}_M\text{e}^B{}_N\,,
\end{equation}
where the second term corresponds to diffeomorphisms along $\xi^M$. This term can be absorbed by choosing the vector $\xi_A=\omega^{-C}{}_{A|C.}$
The left-hand side of the above equation equals
\begin{equation*}
\frac{\text{d}}{\text{d}t}(G_{MN}+B_{MN})= 2\frac{\text{d}\l_{ab}}{\text{d}t}
\left(\text{e}^a{}_M \text{e}^{\hat{b}}{}_N
-\l_{cb}\text{e}^a{}_M \text{e}^c{}_N\right)\,.
\end{equation*}
Employing the above in \eqref{oneloopRG} and reinserting the overall $k_1$, which does not flow, leads to the one-loop $\b$-functions of \eqref{beta}.

\subsubsection{CFT approach}
\label{CFT.connection}

Another approach to the $\b$-functions \eqref{beta} is to employ CFT techniques.
Let us review the results of \cite{LeClair:2001yp}. One considers a perturbation of the form
\be
S_{\rm pert} = \int \text{d}^2\s\sum_A h_A {\cal O}^A\ ,\quad
{\cal O}^A  = \sum_{a,b=1}^{\dim G} d^A_{ab}\, J_+^a J_-^b\ ,
\label{perth}
\ee
where $J^a_+,J^b_-$ satisfy currents algebras at levels $k_1,k_2$ respectively
and $d^A_{ab}$'s are pure numbers that define the perturbation. 
The $d^A_{ab}$'s were
taken to be symmetric in the lower indices $a,b=1,2,\dots,\dim(G)$.
The upper index $A$ takes as many values as the number of independent coupling constants $h_A$.
Making contact with our notation, we have that 
\be
\label{dsosjlfp}
\l_{ab} =  h_A d^A_{ab}\,,
\ee
and so it applies only for symmetric matrices $\l_{ab}$.
The following three conditions ensure closeness of this
algebra and renormalizability at all orders
\begin{equation}
\label{gspdjsl}
d^A_{ab} d^B_{cd} f_{ace} f_{bdf} ={\frak C}^{AB}{}_C d^C_{ef}\ ,\quad
d_{ac}^A d^B_{bc} = {\frak D}^{AB}{}_C d^C_{ab}\ ,\quad d^A_{cd} f_{a e c} f_{e b d} = {\frak R}^A{}_B d^B_{ab}\ ,
\end{equation}
as well as the consistency relations
\begin{equation*}
{\frak C}^{AB}{}_C = {\frak C}^{BA}{}_C\ ,\quad {\frak D}^{AB}{}_C =
{\frak D}^{BA}{}_C\ ,\quad {\frak D}^{AC}{}_D {\frak D}^{DB}{}_E = {\frak D}^{AB}{}_D {\frak D}^{DC}{}_E\ .
\end{equation*}
Finally one defines the quantities
\begin{equation*}
{\frak C}_A(x,y)={\frak C}^{BC}{}_A x_B y_C\ ,\quad {\frak D}^A{}_B = {\frak D}^{AC}{}_B h_C\ ,\quad
\tilde h_A = h_B ((\mathbb{I}-{\frak D}^2)^{-1})^B{}_A  \ .
\end{equation*}
Then the $\beta$-functions are given by \cite{LeClair:2001yp}
\be
\label{dhadt}
\begin{split}
{\mathrm{d}h_A\ov \mathrm{d}t} &=
{1\ov2\sqrt{k_1k_2}} \left(- {\frak C}_B(\tilde h,\tilde h) (\mathbb{I} + {\frak D}^2)^B{}_A\right.\\
&\left.+\left(\l_0+\l_0^{-1}\right)
\left({\frak C}_B(\tilde h{\frak D},\tilde h{\frak D}) {\frak D}^B{}_A  - \tilde h_B ({\frak D}{\frak R}{\frak D})^B{}_A\right) \right)\ ,
\end{split}
\ee
where: $(\tilde h{\frak D})_A = \tilde h_B {\frak D}^B{}_A$.

In fact \eqref{beta} is equivalent to \eqref{dhadt}. Indeed, first note the relation
\begin{equation*}
{\frak D}^A{}_Bd^B_{ab}=d^A_{ac}\l_{cb}\,,
\end{equation*}
where we have used \eqref{dsosjlfp} and the second of \eqref{gspdjsl}. Similarly we find
that
\begin{equation*}
\tilde h_A d^A_{ab}=\l_{ac} g^{cb}\,,
\end{equation*}
where the matrix $g=\mathbb{I}_\text{d}-\l^2$ was defined in \eqref{fospfks}.\footnote{For symmetric coupling matrices $\l_{ab}$, there is no distinction 
between the $g$ and $\tilde g$ defined in \eqref{fospfks}.} Using the above expressions we can easily prove that
\eqref{beta} is equivalent with \eqref{dhadt}, after we contract the latter with $d^A_{ab}$. The terms appearing in 
\eqref{dhadt} are mapped in order to the quadratic, quartic and cubic in $\l_{ab}$'s of \eqref{beta}.

\subsection{The two coupling matrices}
\label{doubleRG}

Before closing this section, we tackle the general case for the two coupling matrices $\l_{1,2}$ for which the action is \eqref{initialaction}.
To compute the one-loop RG flows, one may follow the one-loop effective action approach
or employ gravitational techniques, as in sections \ref{Appadu.general} and \ref{RG.gravity} respectively. However, it is apparent that
the one-loop effective action approach is much simpler. First we consider the action   \cite{Georgiou:2017jfi}
\begin{eqnarray}
\label{initialactionS}
S_{\l_1,\l_2}&=& S_{k_1}(\frak{g}_1) + S_{k_2}(\frak{g}_2)
-{\sqrt{k_1 k_2}\ov \pi} \int \text{d}^2\s\ \Tr\big(A_+ \l_1^{-1} A_-  +B_+ \l_2^{-1} B_- \big)\nonumber
\\
&&+{k_1\ov \pi} \int \text{d}^2\s \ \Tr \big(A_- \del_+ \frak{g}_1 \frak{g}_1^{-1}   - B_+ \frak{g}_1^{-1} \del_- \frak{g}_1+ A_- \frak{g}_1 B_+ \frak{g}_1^{-1}\big) \\
&&+{k_2\ov \pi} \int \text{d}^2\s \ \Tr \big(B_- \del_+ \frak{g}_2 \frak{g}_2^{-1}   - A_+ \frak{g}_2^{-1} \del_- \frak{g}_2+ B_- \frak{g}_2 A_+ \frak{g}_2^{-1} \big) \ .\nonumber
\end{eqnarray}
which after solving for the gauge fields gives \eqref{initialaction}.

The equations of motion of \eqref{initialactionS} are simply {\it two copies} of \eqref{generaleom} for the coupling matrices $\l_1$ and $\l_2$ respectively \cite{Georgiou:2017jfi}
\begin{equation}
\begin{split}
&\l_0\partial_+A_--\l_1^{-T}\partial_-A_+=[\l_1^{-T}A_+,A_-]\,,\\
&\l_1^{-1}\partial_+A_--\l_0^{-1}\partial_-A_+=[A_+,\l_1^{-1}A_-]\, 
\end{split}
\end{equation}
and
\begin{equation}
\begin{split}
&\l_0^{-1}\partial_+B_--\l_2^{-T}\partial_-B_+=[\l_2^{-T}B_+,B_-]\,,\\
&\l_2^{-1}\partial_+B_--\l_0\partial_-B_+=[B_+,\l_2^{-1}B_-]\,.
\end{split}
\end{equation}
The expressions of the gauge fields in terms of the $J^a_{1\pm}$ and the group elements 
is much more complicated than that for the single coupling case in \eqn{gaugefields}. 
These can be
found in \cite{Georgiou:2017jfi} and will not be needed for our purposes.  
Therefore the non-vanishing logarithmic divergent piece, analogs of \eqref{effectiveoneloop} and \eqref{jfklsnckskl} factorizes and
upon integrating over $\text{d}^2p$, the end result simply reads
\be
\begin{split}
-\frac{1}{\pi}\ln\mu&\left(\mathcal{N}_{ac}{}^d(\l_1,\l_0^{-1})\mathcal{N}_{bd}{}^{c}(\l_1^T,\l_0)\big(\l_1^{-1}\big)_{ea}\big(\l_1^{-1}\big)_{fb}A_+^eA_-^f\right.\\
&\left.+\mathcal{N}_{ac}{}^d(\l_2,\l_0)\mathcal{N}_{bd}{}^{c}(\l_2^T,\l_0^{-1})\big(\l_2^{-1}\big)_{ea}\big(\l_2^{-1}\big)_{fb}B_+^eB_-^f\right)\,,
\end{split}
\ee
where the ${\cal N}$'s were given in \eqref{calN}.
Then one can prove on the nose that \eqref{initialactionS} satisfies
\begin{equation}
\label{ofjslshf}
\frac{\text{d}{\cal L}^{(0)}}{\text{d}t}=-\frac{\sqrt{k_1k_2}}{\pi}\left(\frac{\text{d}\big(\l_1^{-1}\big)_{ef}}{\text{d}t}A^e_+A^f_-+
\frac{\text{d}\big(\l_2^{-1}\big)_{ef}}{\text{d}t}B^e_+B^f_-\right)\,.
\end{equation}
Demanding that the effective action is independent of the cutoff scale $\mu$, leads to
\begin{equation}
\label{betadouble}
\boxed{
\begin{split}
&\frac{\text{d}\left(\l_1\right)_{ab}}{\text{d}t}=\frac{1}{2\sqrt{k_1k_2}}\mathcal{N}_{ac}{}^d(\l_1,\l_0^{-1})\mathcal{N}_{bd}{}^{c}(\l_1^T,\l_0)\,,\\
&\frac{\text{d}\left(\l_2\right)_{ab}}{\text{d}t}=\frac{1}{2\sqrt{k_1k_2}}\mathcal{N}_{ac}{}^d(\l_2,\l_0)\mathcal{N}_{bd}{}^{c}(\l_2^T,\l_0^{-1})\,.
\end{split}
}
\end{equation} 
As in the single coupling case, the system is invariant under, implemented as 
in footnote \ref{analyticks}
\begin{equation*}
\l_{1,2}\to\l_{1,2}^{-1}\ , \quad  k_{1,2}\to-k_{2,1} \ 
\end{equation*}
and
\begin{equation*}
\l_{1,2}\to\l_{1,2}^T\ ,\quad k_{1,2}\to k_{2,1}\ ,
\end{equation*}
where the levels $k_{1,2}$ retain their topological nature at one-loop in the large $k_{1,2}$ expansion. 
This set of RG flows is invariant under the interchange of the indices 1 and 2.
For isotropic coupling matrices $\left(\l_{1,2}\right)_{ab}=\l_{1,2}\d_{ab}$ it is in agreement with \eqref{dukjdodlk}.

\section{An application}
\label{RGflows.example}

In this section we analyze the $\b$-function \eqref{beta} of the action \eqref{singlecase} in a simple example, that is a two
coupling case using a splitting of the group indices into subgroup and 
corresponding non-symmetric coset space having special properties.

\subsection{Two coupling case}

Let's split group indices into subgroup $H$ coset $G/H$ indices. For our purposed we will use upper case Latin letters to denote group indices.
We reserve for the subgroup and coset indices lower Latin and Greek letters, respectively.
Consider the case in which the matrix $\l_{AB}$ has elements
\be
\l_{ab} = \l_H \d_{ab}\ ,\quad \l_{\a\b} = \l \d_{\a\b}\ .
\label{traaa}
\ee
Next we compute from \eqref{calN} that
\be
\begin{split}
& \cN_{ab}{}^c(\l;\l_0) = - {\l_H(\l_0-\l_H)\ov 1-\l_H^2} f_{abc}\ ,\quad
\cN_{\a\b}{}^c(\l;\l_0) = {\l^2-\l_0\l_H\ov 1-\l_H^2} f_{\a\b c}\ ,
\\
&
\cN_{\a\b}{}^\g(\l;\l_0) = -{\l(\l_0-\l)\ov 1-\l^2} f_{\a\b\g}\ ,
\quad  \cN_{\a b}{}^\g(\l;\l_0) = -{\l(\l_0-\l_H) \ov 1-\l^2} f_{\a b\g}\ ,
\\
&
\cN_{a\b}{}^\g(\l;\l_0) = -{\l(\l_0-\l_H) \ov 1-\l^2} f_{a\b\g}\ ,
\\
&
\cN_{a\b}{}^c = \cN_{ab}{}^\g = \cN_{\a b}{}^c= 0 \ .
\end{split}
\ee
Then we use the fact that for any semi-simple group $G$
\be
f_{ACD} f_{BCD} = c_G \d_{AB}\ ,\quad f_{acd} f_{bcd} = c_H \d_{ab}\,,\quad
f_{a\g \d} f_{b\g \d} = (c_{G}-c_{H})\d_{ab}\ 
\label{casimii}
\ee
and in addition we assume that
\be
f_{\a\g\d}f_{\b\g\d}=c_{G/H} \d_{\a\b}\ .
\label{dfhji1}
\ee
Unlike \eqn{casimii}, this is not an identity and it holds only for symmetric spaces, where $c_{G/H}=0$, and
for non-symmetric Einstein spaces for which $c_{G/H}\neq 0$.\footnote{The Ricci tensor for a non-symmetric space
with Killing metric $\d_{\a\b}$, reads
\begin{equation*}
R_{\a\b}=f_{\a b\g}f_{\b b\g}+\frac14\,f_{\a\g\d}f_{\b\g\d}=
\frac{c_G}{2}\d_{\a\b}-\frac14\,f_{\a\g\d}f_{\b\g\d}=
\frac{c_G}{4}+\frac12\,f_{\a b\g}f_{\b b\g}\,.
\end{equation*}
Therefore demanding to be an Einstein space, yields \eqref{dfhji1} or \eqref{casim}.
}
 Then it follows that
\be
f_{\a\g c} f_{\b\g c} = \ha (c_{G} -c_{G/H})  \d_{\a\b}\ .
\label{casim}
\ee
One may find non-trivial examples for which \eqn{dfhji1} holds with non-vanishing right hand side.
In particular, in investigations of ten-dimensional compactifications of gravity backgrounds and of gauge theories to four dimensions, the following three non-trivial six dimensional examples have been encountered  \cite{Forgacs:1985vp,Lust:1986ix,Mueller,Manousselis:2000aj}, listed in table \ref{tablecosets}.
\begin{table}
\begin{center}
  \begin{tabular}{ | c | c | c | c|}
  \hline
   \text{Cosets} & $c_G$ & $c_H$ & $c_{G/H}$\\ \hline
    \hline
    $SU(3)/U(1)^2$ & 6 & 0 & 2 \\ \hline
    $Sp(4)/SU(2)\times U(1)$ & 4 & 4 & 2 \\ \hline
    $G_2/SU(3)$ & 8 & 6  & 8/3\\ \hline
  \end{tabular}
  \caption{Non-trivial six dimensional examples of non-symmetric Einstein spaces.} \label{tablecosets}
\end{center}
\end{table}
In general $c_G>c_H,c_{G/H}$ but there is conclusion for the relation between 
$c_H$ and $c_{G/H}$.

It turns out that the truncation \eqn{traaa} is a consistent if and only if
\eqn{dfhji1} is satisfied
\be
\label{RGcosetn}
\begin{split}
&
{\mathrm{d}\l_{H} \ov \mathrm{d}t} = -{(\l_H-\l_0)(\l_H-\l_0^{-1})\ov 2\sqrt{k_1k_2}}
\left( c_H {\l^2_H \ov (1-\l_H^2)^2} +
(c_G-c_H) {\l^2 \ov (1-\l^2)^2} \right)\ ,
\\
&
{\mathrm{d}\l \ov \mathrm{d}t} =
 -{1\ov 2\sqrt{k_1k_2}} \left(c_{G/H} {\l^2 (\l-\l_0)(\l-\l_0^{-1})\ov (1-\l^2)^2}
 + {c_G-c_{G/H}\ov 2}\right.
 \\
 & \left. \times {\l \ov (1-\l^2) (1-\l_H^2)}
\left((\l_0^{-1}-\l_H)(\l_0\l_H-\l^2)+ (\l_0-\l_H)(\l_0^{-1}\l_H-\l^2)\right) \right)\ ,
\end{split}
\ee
which is invariant under
\be
\l\to\l^{-1}\ , \quad \l_H\to\l_H^{-1} \ ,\quad  k_{1,2}\to-k_{2,1} \ .
\label{dhi2}
\ee
This symmetry if inherited by corresponding symmetry for the $\s$-models backgrounds.
In implementing it one should treat it as an analytic continuation when square roots appear, 
as in footnote \ref{analyticks}.
Hence, $(k_1, k_2) \to \text{e}^{i\pi} (k_2,k_1)$ and $(1-\l) \to \text{e}^{-i \pi }\l^{-1}(1 -\l)$.

\no
The above system of RG flow equations has the fixed points
\begin{equation*}
(\l_H,\l)=(0,0),(\l_0,0),(\l_0,\l_0),(\l_0^{-1},0),(\l^{-1}_0,\l^{-1}_0)\ .
\end{equation*}
We may assume without loss of generality that $0<\l_0<1$. From this and the fact that the levels are positive, we deduce that the physical ones are the first three.
The RG flow using those fixed points are depicted in the Fig. 1. 
\begin{figure}
\label{betaplot}
\begin{center}
\vskip -0 cm
\includegraphics[height= 8 cm,angle=0]{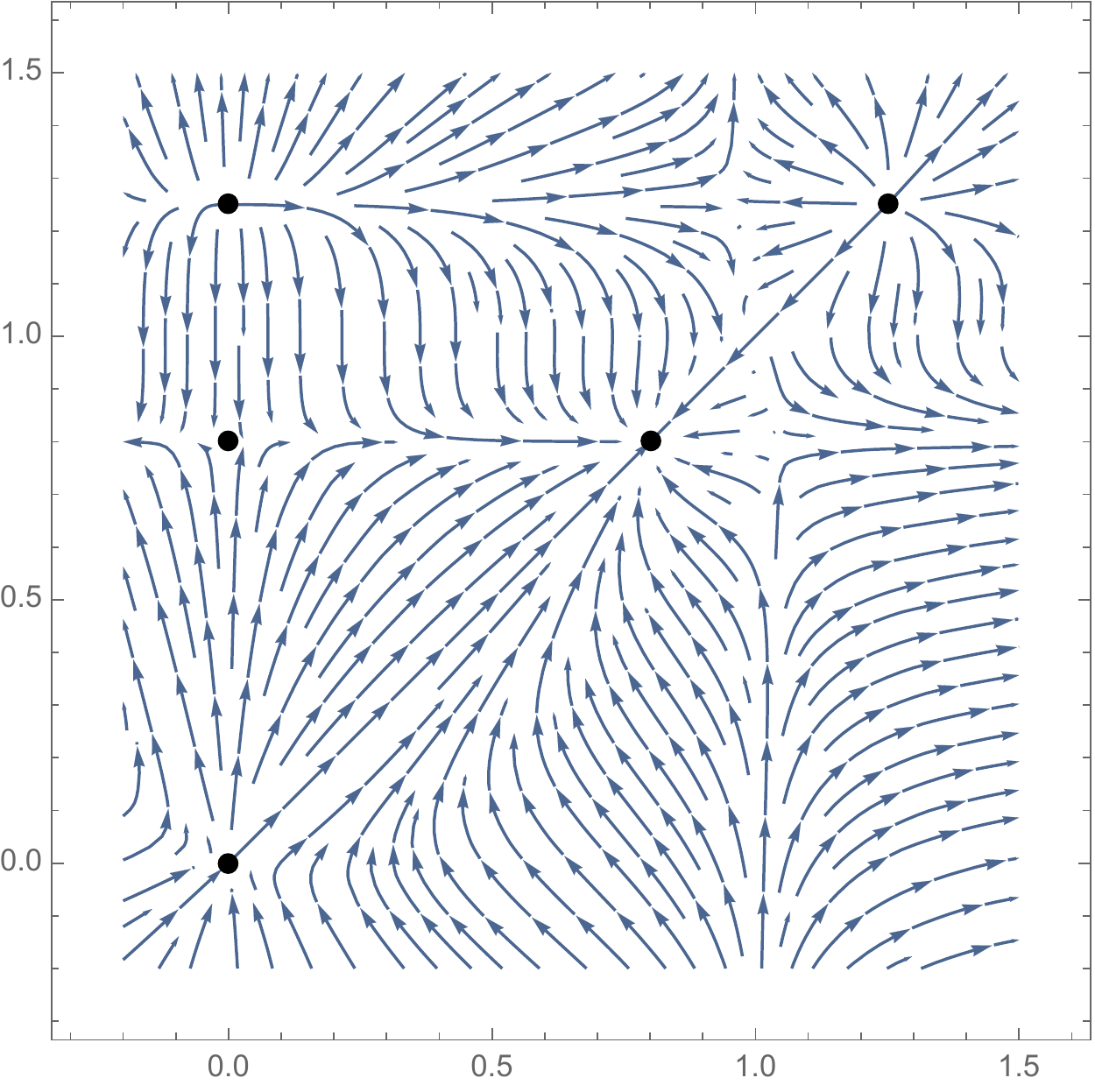}
\end{center}
\vskip -0.7 cm
\caption{RG flows in the $(\l,\l_H)$ plane ($\l$-horizontal) for the $G_2/SU(3)$ coset, see Table \ref{tablecosets} with $\l_0=0.8$.
Clearly the point $(\l_0,\l_0)$ is an IR attractor.}
\label{fig:coupled}
\end{figure}

\no
Setting $\l_H=\l_0$ is consistent with the corresponding equation of motion in \eqref{RGcosetn},
provided that $\l_0\neq1$. Then, for the remaining coupling we find that
\be
\label{betacosetn}
{\mathrm{d}\l \ov \mathrm{d}t} = -{c_G-c_{G/H}\ov 4k_1} {\l (\l_0^2-\l^2)
\ov 1-\l^2}
- {c_{G/H} \ov 2 \sqrt{k_1 k_2}}{ \l^2(\l_0-\l)(\l_0^{-1}-\l) \ov (1-\l^2)^2} \ .
\ee
This $\b$-function has the following properties:
\begin{enumerate}
\item
It is invariant under the symmetry $\l\to\l^{-1}$ and $k_{1,2}\to-k_{2,1}$, which is the left over symmetry \eqn{dhi2} after setting $\l_H=\l_0$.

\item
It has a fixed point at $\l=0$, near which the $\b$-function \eqref{betacosetn} reads
\begin{equation*}
\b\simeq-\frac{(c_G-c_{G/H})\l}{4k_2}+{\cal O}(\l^2)\,.
\end{equation*}
Hence, the operator that drives the perturbation is relevant and has dimension
\begin{equation*}
\Delta=2-\frac{c_G-c_{G/H}}{2k_2}\,.
\end{equation*}
It is not clear to us whether or not  this fixed point corresponds to an exact CFT.

\item
It has an IR fixed point at $\l=\l_0$, provided that $\l_0\neq 1$. At that point the CFT 
corresponding to the action \eqn{singlecase} has been identified in $G_{k_1}\times G_{k_2-k_1}$ CFTs \cite{Georgiou:2017jfi}.
The $\b$-function \eqref{betacosetn} for $\l$ near $\l_0$ reads
\begin{equation*}
\b\simeq\frac{c_G(\l-\l_0)}{2(k_2-k_1)}+{\cal O}(\l-\l_0)^2\,,
\end{equation*}
and so the operator that drives the perturbations has dimension
\begin{equation*}
\Delta=2+\frac{c_G}{k_2-k_1}\,.
\end{equation*}

\item
For equal levels $\l_0=1$,  the $\b$-function \eqref{betacosetn} reads
\be
\label{betacosett}
{\mathrm{d}\l \ov \mathrm{d}t} = -{1\ov 2k} \left(  \frac{c_G-c_{G/H}}{2} \l  +c_{G/H} {\l^2 \ov (1+\l)^2} \right)\ .
\ee
Note however, that setting $\l_H=1$ is not a fixed point of the flow \eqref{RGcosetn}, except if the subgroup $H$ is an Abelian one.
Nevertheless the expression \eqref{betacosett} remains valid as we prove in
Section \ref{Coset.Appadu} through a direct computation in the coset.
This $\beta$-function \label{betacoset} does not admit a new fixed point in the IR. Instead,
the theory is driven at strong coupling.

\end{enumerate}

\subsection{Coset computation}
\label{Coset.Appadu}

We will compute the $\beta$-function \eqn{betacosett} using  the one-loop effective action projected at the coset $G/H$.
We need to determine a specific background solution
and evaluate its quantum fluctuations, following as before the techniques presented in
\cite{Appadu:2015nfa,Georgiou:2017aei}.
Specializing the equations of motion for the subgroup and coset $(A^h_\pm=A^a_\pm t_a,A^{g/h}_\pm=A_\pm^\a t_\a)$
and the coupling matrix $\l_{AB}$
\begin{equation*}
\l_{ab}=\delta_{ab}\,,\quad \l_{\a\b}=\l\delta_{\a\b}\,,
\end{equation*}
we find that \cite{Georgiou:2016urf}
\begin{equation}
\label{eomApmcoset}
\begin{split}
&\partial_\pm A^{g/h}_\mp=-[A^{g/h}_\mp,A^h_\pm]\pm\a[A^{g/h}_+,A^{g/h}_-]\,,\\
&\partial_+A^h_--\partial_-A^h_+=[A^h_+,A^h_-]+\b[A^{g/h}_+,A^{g/h}_-]\,,\\
&\a=\frac{1}{1+\l}\,,\quad \b=\frac{1}{\l}\,.
\end{split}
\end{equation}
Moreover we fix the residual gauge through the covariant gauge fixing condition
\begin{equation}
\label{cosetgauge}
\partial_+A^h_- +\partial_-A^h_+=0\,.
\end{equation}
At first we specify a background solution of \eqref{eomApmcoset} and \eqref{cosetgauge}
\begin{equation*}
A^a_\pm=0\,,\quad A^\a_+=i\l\ J^{\a}_{1+}\,,\quad A^\a_-=-i\l J^{\a}_{2-}\,,
\end{equation*}
where we set $A^h_{\pm}=0$, so that we project to the coset $G/H$.
The Lagrangian density for this background reads
\begin{equation}
\label{clasLcoset}
{\mathcal L}^{(0)}={\mathcal L}_{k_1}(\frak{g}_1)+{\mathcal L}_{k_2}(\frak{g}_2)
+\frac{k}{\pi}\, \l\, J_{1+}^{\a} J^{\a}_{2-}\,.
\end{equation}
Next we vary the equations of motion \eqref{eomApmcoset} and the covariant gauge fixing
condition \eqref{cosetgauge} obtaining for the fluctuations $(\d A^{g/h}_\pm,\d A^h_\pm)$
\begin{equation}
\left(\begin{array}{cccc}
\partial_- +\a \tilde{A}^{g/h}_{-}& -\a \tilde{A}^{g/h}_{+} & 0 & -\tilde{A}^{g/h}_{+}	\\
-\a \tilde{A}^{g/h}_{-} & \partial_+ +\a \tilde{A}^{g/h}_{+} & -\tilde{A}^{g/h}_{-} & 0	\\
-\b \tilde{A}^{g/h}_{-} & \b \tilde{A}^{g/h}_{+} & -\partial_- & \partial_+	\\
0 & 0 & \partial_- & \partial_+
\end{array}\right)
\left(\begin{array}{c}
\d A^{g/h}_{+}\\
\d A^{g/h}_{-}\\
\d A^h_{+}\\
\d A^h_{-}
\end{array}\right)=0\,,
\end{equation}
with
$\left(\tilde{A}^{g/h}_{\pm}\right)_{AB}=i f_{AB\g}\, A^\g_{\pm}$.
To evaluate the one-loop effective Lagrangian, we Wick rotate to Euclidean space and then
we integrate out the fluctuations
in the Gaussian path integral. The result in momentum space reads
\begin{equation}
-{\mathcal L}^\text{eff}_\text{E}={\mathcal L}^{(0)}+
\int^\mu\frac{\text{d}p_1\text{d}p_2}{(2\pi)^2}\ln\det{\cal D}^{-1/2}\,,
\end{equation}
where
\begin{equation}
{\cal D}=\left(\begin{array}{cccc}
p_- +\a \tilde{A}^{g/h}_{-}& -\a \tilde{A}^{g/h}_{+} & 0 & -\tilde{A}^{g/h}_{+}	\\
-\a \tilde{A}^{g/h}_{-} & p_+ +\a \tilde{A}^{g/h}_{+} & -\tilde{A}^{g/h}_{-} & 0	\\
-\b \tilde{A}^{g/h}_{-} & \b \tilde{A}^{g/h}_{+} & -p_- & p_+	\\
0 & 0 & p_- & p_+
\end{array}\right)\,.
\end{equation}
After some algebra we find that
\begin{equation}
\begin{split}
\label{effectiveoneloopcoset}
-{\mathcal L}_E^\text{eff}&={\mathcal L}^{(0)}+\frac{\ln\mu}{\pi}\,
\l^2\left(\alpha^2f_{\a\b\g}f_{\a\b\d}+\b f_{\a b\g}f_{\a b\d}\right)J_+^{1\g}J_-^{2\d}\\
&=\cdots+\frac{1}{\pi}\left( k\l\,\d_{\g\d}+\frac{\ln\mu}{\pi}\,
\l^2\left(\alpha^2f_{\a\b\g}f_{\a\b\d}+\b f_{\a b\g}f_{\a b\d}\right) \right)J^{1\g}_{+}J^{2\d}_{-}\,,
\end{split}
\end{equation}
where the dots denote as before $\l$-independent terms.
Again, the one-loop RG flows can be found by demanding that \eqref{effectiveoneloopcoset} is independent
of the cutoff scale $\mu$. Using \eqref{dfhji1}, \eqref{casim} and the definitions of $\a,\b$ in \eqref{eomApmcoset} we obtain \eqref{betacosett}.

\section{Outlook}
\label{concl}

We studied quantum properties of the actions \eqref{initialaction} and 
\eqn{singlecase} describing smooth interpolations between exact CFTs \cite{Georgiou:2017jfi}. We proved that are one-loop renormalizable and 
we derived its renormalization group flows for general coupling matrices in 
\eqn{betadouble} and \eqn{beta}. The derivation was achieved by computing the one-loop effective action of fluctuations around a background solution and from gravitational techniques. Our results are in agreement with limit cases existing in the literature. Namely, for symmetric couplings and different levels  in \cite{LeClair:2001yp} and for general couplings but equal levels in \cite{Sfetsos:2014jfa}. 
We elucidated our results by studying the two coupling case arising form splitting the group  into a subgroup and  the corresponding coset manifold. This is consistent if the latter is either a symmetric-space or a non-symmetric Einstein-space. 
It is interesting to compute correlation functions and anomalous dimensions of operators in these two-coupling theories. This would generalize analogous computations for the isotropic case in \cite{Georgiou:2015nka,Georgiou:2016iom,Georgiou:2016zyo}. 
 Finally it would interesting to apply the one-loop effective action techniques for the closely related $\eta$-deformed models which were introduced
 for semi-simple groups and symmetric cosets in \cite{Klimcik:2002zj,Klimcik:2008eq,Delduc:2014uaa} and \cite{Delduc:2013fga,Delduc:2013qra} respectively.
 The goal would be to derive the general one-loop RG flow equations as has been performed only for
isolated cases \cite{Squellari:2014jfa,Sfetsos:2015nya,Demulder:2017zhz}.

\subsection*{Acknowledgements}

We acknowledge each others home institutions for warm hospitality and financial support.


\end{document}